\documentclass[reprint,notitlepage,onecolumn,superscriptaddress,amsmath,amssymb]{revtex4-1}
\usepackage[utf8]{inputenc}
\usepackage{graphicx}
\usepackage{diagbox}
\usepackage{braket}
\usepackage{subfigure}
\usepackage{multirow}
\usepackage{tabularx}
\usepackage{booktabs}
\usepackage{todonotes}

\begin{document}
	
\title{An efficient combination strategy for hybird quantum ensemble classifier}
\author{Xiao-Ying Zhang}
\author{Ming-Ming Wang}
\email{bluess1982@126.com}
\affiliation{Shaanxi Key Laboratory of Clothing Intelligence, School of Computer Science, Xi'an Polytechnic University, Xi'an 710048, China}

\begin{abstract}
Quantum machine learning has shown advantages in many ways compared to classical machine learning. In machine learning, a difficult problem is how to learn a model with high robustness and strong generalization ability from a limited feature space. Combining multiple models as base learners, ensemble learning (EL) can effectively improve the accuracy, generalization ability, and robustness of the final model. The key to EL lies in two aspects, the performance of base learners and the choice of the combination strategy. Recently, quantum EL (QEL) has been studied. However, existing combination strategies in QEL are inadequate in considering the accuracy and variance among base learners. This paper presents a hybrid EL framework that combines quantum and classical advantages. More importantly, we propose an efficient combination strategy for improving the accuracy of classification in the framework. We verify the feasibility and efficiency of our framework and strategy by using the MNIST dataset. Simulation results show that the hybrid EL framework with our combination strategy not only has a higher accuracy and lower variance than the single model without the ensemble, but also has a better accuracy than the majority voting and the weighted voting strategies in most cases.
\end{abstract}

\date{today}

\maketitle

\section{Introduction}

Based on the basic principles of quantum mechanics, quantum computing provides new models for accelerating solutions of some classical problems \cite{Shor97,Grover97,HHL-2009}. With the great success of machine learning \cite{LecunBengio-313,GoodfellowBengio-312}, quantum machine learning \cite{abohashima2020classification,biamonte2017quantum} have been developed to show characteristics of quantum acceleration.
Typical examples include quantum neural networks (QNNs) \cite{Kak-174,Ronald-173}, quantum principal component analysis \cite{lloyd2014quantum}, quantum support vector machine \cite{rebentrost2014quantum}, quantum unsupervised learning \cite{wiebe2014quantum}, quantum linear system algorithm for dense matrices \cite{wossnig2018quantum}, etc.

Neural networks are at the center of machine learning. As its counterpart, a variety types of QNN models have been proposed since its first appearance \cite{Kak-174,Ronald-173}, which include models based on quantum dot \cite{BehrmanNiemel-189}, on superposition \cite{VenturaMartinez-190}, on quantum gate circuits \cite{MatsuiTakai-192}, on quantum-walk \cite{SchuldSinayskiy-194}, and quantum analogue of classical neurons \cite{BeerBondarenko-166}, etc. In the last few years, quantum deep learning \cite{WiebeKapoor-115}, quantum convolutional neural networks (QCNNs) \cite{CongChoi-181,kerenidis2019quantum,hur2022quantum}, quantum generative adversarial network \cite{LloydWeedbrook-180}, quantum autoencoders \cite{RomeroOlson-104}, have also been developed. Most of QNNs are constructed by parameterized quantum circuit (PQC) \cite{plesch2011quantum,schumacher1995quantum}. In a PQC model, the number of parameters, the types of quantum gates, and the width and depth of the circuit have a great impact on the required resources, the difficulty of solving gradients, and whether the optimal model can be obtained \cite{du1810expressive,sim2019expressibility}.

Ensemble learning (EL), also known as the multi-classifier system, is an important algorithm that combines multiple learning models to achieve better performance. In 1990, Schapire pointed out that it is possible to surpass one strong learner by combining several weak learners (base learners) in a roughly correct sense \cite{schapire1990strength}. It lays the foundation for the Adaboost algorithm \cite{sagi2018ensemble}. EL  has shown advantages in avoiding over-fitting, reducing the risk of decision error, and reducing the acquisition of local minimum \cite{sagi2018ensemble}. It has been widely used in object detection \cite{paisitkriangkrai2017structured}, education \cite{beemer2018ensemble}, malware detection \cite{idrees2017pindroid}, etc.
The performance of a EL mainly depends on the diversity and the prediction performance of base learners. The diversity of the EL can be realized by using different model structures, training sets, method of subsetting, and others \cite{sagi2018ensemble}. While the prediction performance of the EL is correlated with the uncorrelated degree of error among base learners \cite{ali1995link}.
An important part of a EL is the combination strategy for combinating the predictions of base learners. Currently, combination strategies can be roughly divided into weight combination and meta-learning methods. In terms of a EL in binary classification tasks, the classification accuracy of each base learner must be better than random guessing, such that the EL can be effective. In addition, diversity among models should still be maintained. There are many EL methods related to classical learning, such as AdaBoost, bagging, stacking, random forest, and so on \cite{quinlan1996bagging,wwys2018}.

For quantum computing, some studies have been performed on quantum ensemble learning (QEL) \cite{schuld2018quantum,macaluso2020quantum,macaluso2020quantum1,araujo2020quantum}. In Ref. \cite{schuld2018quantum}, Schuld et al. proposed a framework to construct ensembles of quantum classifiers, which evaluate the predictions of exponentially large ensembles with parallelism.
In Refs. \cite{macaluso2020quantum,macaluso2020quantum1}, a QEL scheme using the bagging was proposed by using quantum cosine classifier as base learners, and numerical simulation experiments were carried out. However, their quantum base learner is untrained. The implementation of the classifier is just for one test at one time, which is not suitable for practical applications. Based on Ref. \cite{schuld2018quantum}, Araujo et al. \cite{araujo2020quantum} further proposed the QEL of trained classifiers, which shown the advantage of QEL for quantum classifiers. But the increasing of data size optimization steps has a great impact on the model since it is difficult to execute on a quantum computer. 
Their combination process can not effectively differentiate the difference among base learners, which will impair the accuracy of the result.

In this paper, a quantum-classical hybrid EL model is proposed. That is, quantum classifiers are used as base learners, and the bagging EL method \cite{sun2011bagging} is used to assemble quantum classifiers in a classical computer.
Since the previous QEL framework \cite{araujo2020quantum} can not clearly distinguish the differences among base learners, while classical combination strategies can not distinguish the similarity among base learners, we propose a new combination strategy that considers the difference among base learners and the performance of each learner in our framework.
The MindQuantum platform \cite{mq_2021} is used as the training platform for a single quantum base learner.
The MNIST handwritten digits \cite{lecun1998gradient} are used as the dataset. The feasibility of our strategy is verified by using the homogenous ensemble method that uses the same structural base learner as ensemble members.

\section{Ensemble learning}

EL mainly divided into the ``homogeneous" ensemble and the ``heterogeneous" ensemble, where their main difference lies in whether the models of classifiers adopt the same model structure. A single learner in EL is called a base learner. The key to the final effectiveness of the ensemble lies in whether the characteristic with ``good but different" can be maintained among the base learners. That is, each base learner should have a certain accuracy that is better than random guess and there are certain differences between them \cite{sagi2018ensemble,dong2020survey,xjw2018}. For a binary classification problem, assuming that the accuracy rate of each base classifier is $p$ and the base classifiers are independent of each other, a simple voting method is adopted to combine $N$ base classifiers, then the error rate of the ensemble result is \cite{xjw2018}
	\begin{equation}
		\sum_{k=0}^{N/2}
		\left(\begin{matrix}
			N  \\
			k  \\
		\end{matrix}\right)p^k (1-p)^{N-k}.
		\label{eq1}
	\end{equation}
Under ideal conditions, the error rate will gradually decrease and eventually approach 0 with the increase of the number of base classifiers. Assuming the accuracies of the two classifiers are $acc_1$ and $acc_2$ with $\frac{1}{2} \leq acc_1 \leq acc_2 \leq 1$. The interval of similarity of two classifiers is $[acc_1-(1-acc_2),acc_1+(1-acc_2)]$. The higher the accuracy of the base classifier is, the higher the similarity between models will be, and the smaller the difference will be, in which case the ensemble is likely to be invalid. So it is almost impossible to reach ideal conditions.
	
In addition to the performance requirements among base learners and their models, the combination strategy is also an important factor affecting the results. In general, models need to be pruned before using the combination strategy, i.e., to determine which models can participate in the combination. For example, the sorting or search-based strategies can be used to filter models \cite{sun2011bagging}. The models' predictions are then combined. Commonly used combination strategies include the averaging strategy, voting strategy, and learning strategy \cite{zhang2012ensemble}. Among them, the voting strategy is divided into absolute majority voting, pluraity voting and weighted voting \cite{xjw2018}.
	
\section{Hybrid quantum-classical ensemble learning}

\begin{figure}[ht]
		\centering
		\includegraphics[scale=0.7]{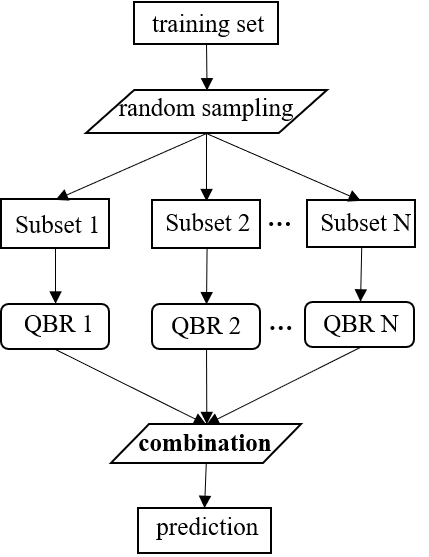}
		\caption{The structure of hybrid quantum-classical EL, where QBR means quantum base learner. }
		\label{fig2}
	\end{figure}

We present a hybrid EL framework that combines quantum computation with classical computation. It takes advantage of the parallelism of quantum computing and the convenience of classical computer processing parameters.
The hybrid learning framework is shown in Fig. \ref{fig2}. Taking the image classification task as an example, the image dataset is firstly reduced in the classical computer to preserve $2^n$ features, where $n$ is the number of qubits used by the quantum base learner. The dataset is divided into a training set and a test set.
By random sampling, $N$ subsets are got from the training set, where $N$ is the number of the quantum base learner. These data are encoded into quantum states before the quantum base learners are trained in a quantum computer. The training of a quantum base learner can be completed in parallel by multiprocessing. Each quantum base learner predicts for the test set and outputs the decision result as the final result on the classical computer according to the combination strategy.
fig3
	
\subsection{Quantum base learner}
For EL of classification tasks, any classifier can be used as a base learner as long as the classifier can achieve classification tasks slightly better than random guess.
To use EL as a large network for training in a practical application, the memory cost of the system is very high. For example, by using classical neural network as the base learner, the convergence rate of each base learner is slow in the training process, even in parallel computing. While QNNs can use fewer quantum resources for achieving a similar results to the classical neural network at a faster convergence rate \cite{hur2022quantum}.
		
\begin{figure}[ht]
\centering
	\includegraphics[scale=0.5]{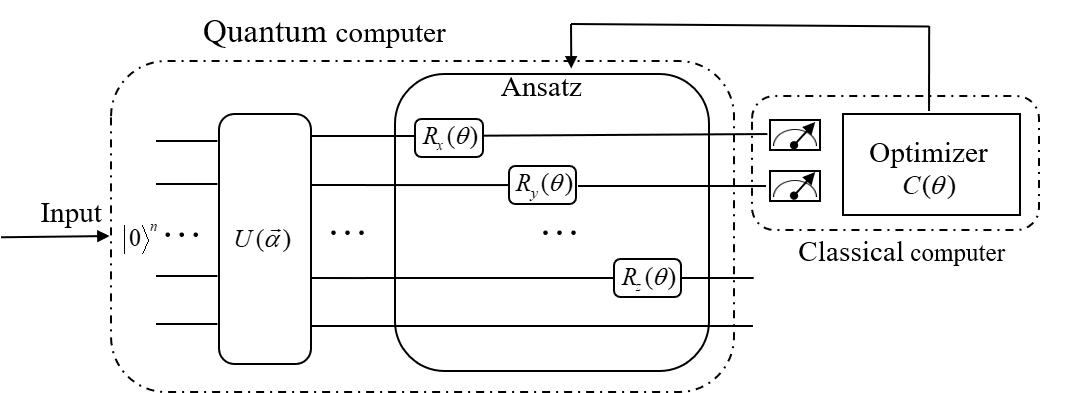}
	\caption{The training model of a quantum base learner.}
	\label{fig1}
\end{figure}

Variational quantum algorithms (VQA) have been developed because of their quantum advantages for noisy intermediate-scale quantum (NISQ) hardware. The design of VQA mainly includes three parts. Firstly, classical data are encoded into quantum states. Secondly, we need to design an Ansatz, a set of continuous or discrete parameters that depend on quantum operators, for training optimization. Thirdly, the loss function is designed according to the objective to determine the optimization direction \cite{cerezo2021variational}. The training model is shown in Fig. \ref{fig1}, where trainable parameters are encoded into the parameterized quantum circuit $U(\vec{\alpha})$ \cite{plesch2011quantum,schumacher1995quantum}. The task is carried out on a quantum computer, while the loss calculation and parameters optimization are performed on a classical computer.
For different task objectives, we use different network structure of Ansatz and loss functions \cite{cerezo2021variational}.
	
\section{The combination strategy}
The diversity among base learners, the training of base learners, and the combination of base learner are three significant factors in EL. Weighted voting is a commonly used method in combination strategies, which is defined as \cite{xjw2018}
\begin{equation}
H(x) = C_{\arg_j\max\sum_{i=1}^{N}w_i h_{i}^{j}(x)},
\label{eq2}
\end{equation}
where $C_{\arg_j}$ is the data label, $h_{i}^{j}(x)$ is the output of the learner on the classes, $w_i$ is the weight of $h_{i}$, usually $w_i \geq 0$. The voting method with classification accuracy as the weight is simple and effective. However, this blindly voting method will lead to the combination result approach to a certain type of base learners if some base learners are the same or have very small difference without pruning for all base learners. 
Instead, it reduces the efficiency of ensemble. To deal with this problem, we propose a weight voting strategy based on the confusion matrix to obtain the similarity among base learners and individual performance as weights.

As an important method to measure the model performance, confusion matrix can be used to calculate recall rate, accuracy rate, and accuracy rate, which could be considered as fixed properties of the model to a kind of dataset. It can be used to improve  ensemble classifiers\cite{marom2010using}. For binary classification, the confusion matrix is shown in Tab. \ref{tab1}, where \textit{TN} is the number of samples predicted to be 0 for test data and true to be 0, \textit{FN} is the number predicted to be 0 and true to be 1, \textit{TP} is the number of samples predicted to be 1 and true to be 1 for test data, and \textit{FN} is the number predicted to be 1 and true to be 0.
\begin{table}[ht]
\caption{The confusion matrix of a binary classification.}	
\label{tab1}
\begin{tabular}{c|cc}
			\hline
			\diagbox{true}{predict} & 0   &   1 \\
			\hline
			0 &  \textit{TN}  &  \textit{FP}  \\
			\hline
			1 &  \textit{FN}  & \textit{TP}  \\
			\hline
\end{tabular}
\end{table}

Our combination strategy is as follows. At the end of the training, the confusion matrix of each learner and the similarities with others are obtained by predictions of data randomly sampled for the training set. In order to efficiently distinguish the voting weight of base learners with low accuracy and high similarity, which acts as model pruning, the value of the highest similarity between the current learner and the later learner is denoted as $S_{\max}$. Then, $-\log(S_{\max})$ is regard as part of the weight coefficient to reduce its voting weight when $S_{\max}> S_{\text{threshold}}$, where $S_{\text{threshold}}$ is the threshold of similarity for base learners. Therefore, when the base learner predicts the new data as 0 and the max similarity exceeds the threshold, the voting weight is defined as
\begin{equation}
	w = acc\times(\frac{TN}{TN+FN}+\frac{TN}{TN+FP})\times(-\log(S_{\max})),
	\label{equation3}
\end{equation}
where $acc$ refers to the accuracy of a base learner. When the learner predicts the data to be 1, the voting weight is defined as
\begin{equation}
w = acc\times(\frac{TP}{TP+FP}+\frac{TP}{TP+FN})\times(-\log(S_{\max})).
\label{equation4}
\end{equation}

The voting weight obtained by the base learner is lower if the model similarity is higher. If lots of base learners with high similarity appeared in the model, the influence of ensemble prediction will be little, which is equivalent to prune the acquired base learners.

Moreover, we set the voting weight as $S_{\max}$ for substituting $-\log(S_{\max})$ in Eqs. (\ref{equation3}) and (\ref{equation4}) when the accuracy of the base learner is bigger than $accuracy\_threshold$. This effectively preserves the ability of base learners with good classification performance in random testing to improve ensemble results.

\section{Numerical simulation}

We use the open boundary QCNN network \cite{hur2022quantum} as our base learners for simulating our hybrid learning framework. As shown in Fig. \ref{fig4}, the quantum network acts as a base learner in Bagging and trains the parameters of the model with the images after dimension reduction with PCA. Finally, the combination and optimization are completed on a classical computer, as shown in Fig. \ref{fig4}.
	
In our simulation, MNIST digits are used as the dataset, which are divided into the training and the test sets.
The accuracy and variance in prediction for the test set are used as the comparison data between our QEL framework and a single learner without EL. The classical data will be converted into quantum data with the amplitude coding. The quantum state $\ket{\phi(x)}$ is encoded as \cite{araujo2021divide}
\begin{equation}
U_\theta (x)\ket{0}^{\otimes n} \rightarrow \ket{\phi(x)}=\frac{1}{\Vert x \Vert}\sum_{i=1}^{n}x_i\ket{i}.
\label{eq3}
\end{equation}

\begin{figure}[ht]
		\centering
		\subfigure[]{\includegraphics[scale = 0.6]{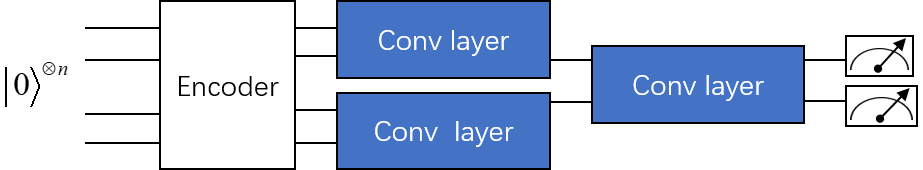}}
		\hspace{1cm}
		\subfigure[]{\includegraphics[scale = 0.6]{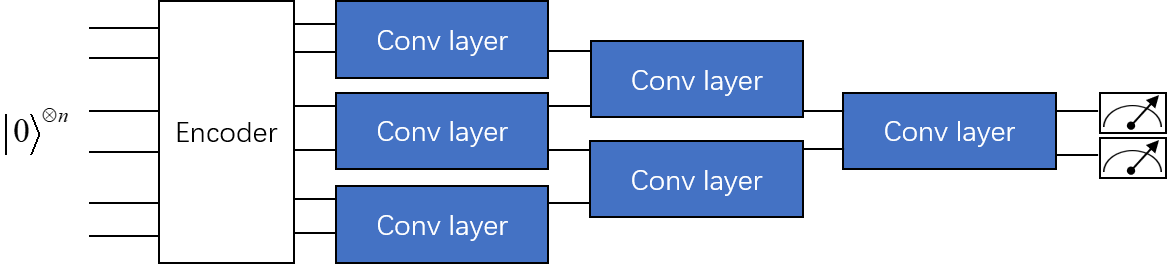}}
		\hspace{1cm}
		\subfigure[]{\includegraphics[scale = 0.6]{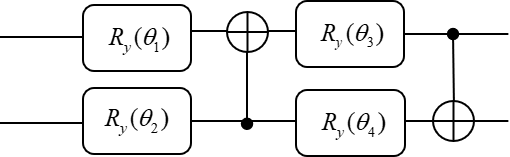}}
		\vspace{0.5cm}
		\subfigure[]{\includegraphics[scale = 0.6]{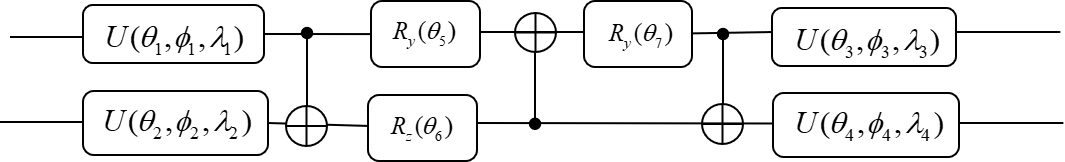}}
		\centering
		\caption{The models and circuits of our quantum base learners. (a) The structure of open boundary QCNN model in Ref. \cite{hur2022quantum} for 4-qubit where Conv layer means the convolutional layer; (b) The structure of QCNN model for 6-qubit in Ref. \cite{hur2022quantum}; (c) The circuit 3 in Ref. \cite{hur2022quantum}; (d) The circuit 9 in Ref. \cite{hur2022quantum}.}
		\label{fig3}
\end{figure}

\begin{figure}[ht]
		\centering
		\includegraphics[scale=0.7]{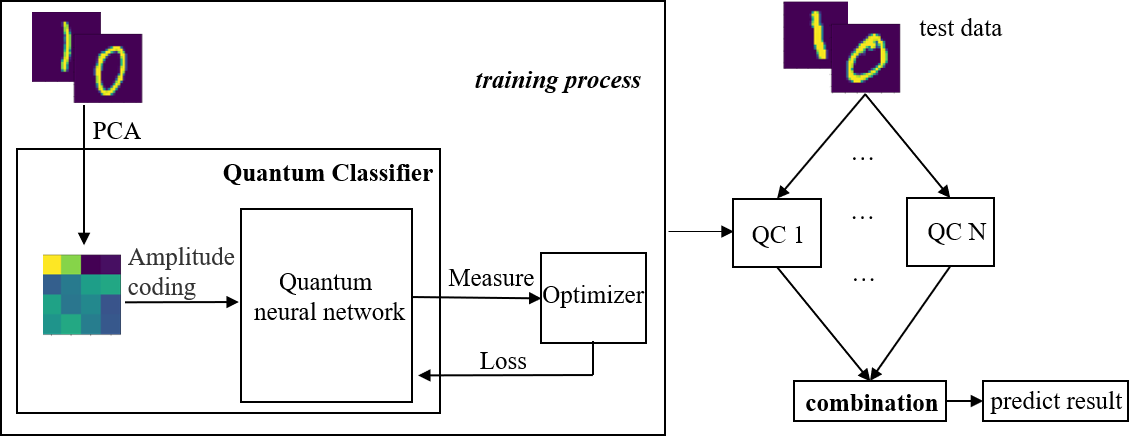}
		\caption{The process our QEL framework, where QC means quantum classifier}
		\label{fig4}
\end{figure}

Convolutional circuit 3 and 9 in Ref. \cite{hur2022quantum}, which are shown in Fig. \ref{fig3} (c) and (d), are used as convolution layer of base learners in Fig. \ref{fig3} (a) and (b), respectively.
The dimensionality of training data is reduced with PCA firstly, then they are encoded into 4-qubit states by the amplitude coding.
Two types of base learners are used here, i.e., the 4-qubits and 6-qubits base learners.
Our 4-qubits base learners are QCNNs shown in Fig. \ref{fig3} (a) by using the circuit in Fig. \ref{fig3} (c) as convolutional layers, while our 6-qubits base learners are QCNNs shown in Fig. \ref{fig3} (b) by using the circuit in Fig. \ref{fig3} (d) as convolutional layers.

\begin{figure}[ht]
		\centering
		\subfigure[]{\includegraphics[scale = 0.35]{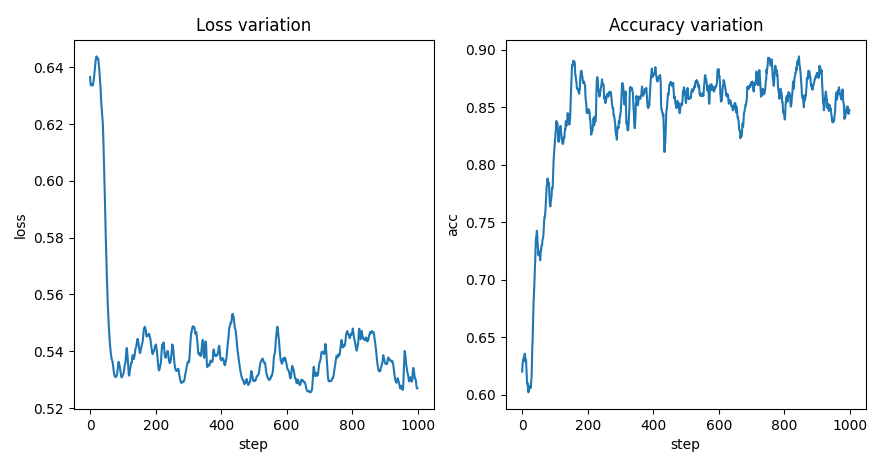}}
		\hspace{0.00cm}
		\subfigure[]{\includegraphics[scale = 0.35]{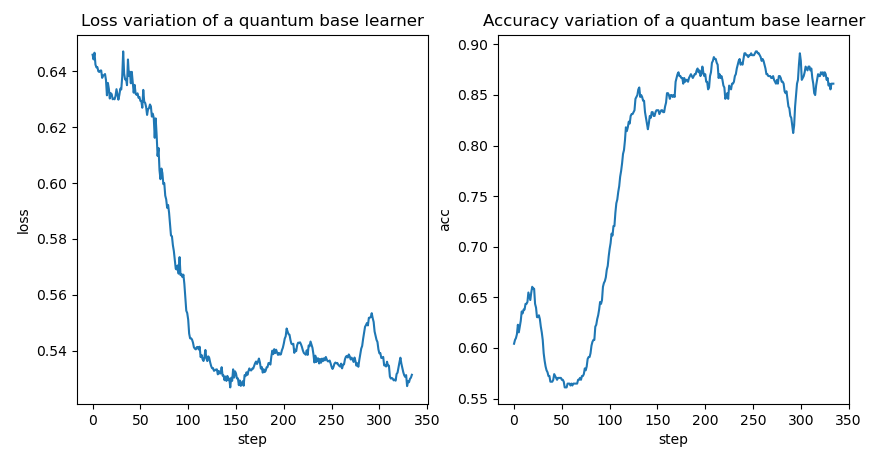}}
		\centering
\caption{The comparison of variation of loss and accuracy. (a) The variation of loss and accuracy of the single learner without ensemble; (b) The variation of loss and accuracy of quantum base learners.}
\label{fig5}
\end{figure}

We compared the variation of loss and accuracy between a single learner which trained the whole dataset without ensemble
and a set of base learners which trained with a training subset getting by randomly sampling.
As is shown in Fig. \ref{fig5}, the quantum base classifier can achieve approximate loss and accuracy compared with a single model in a shorter steps. This shows that the quantum classifier can classify the test set well even if the model is trained with partial data. It is feasible to train quantum base classifiers through partial data and combine them with an effective combination strategy to achieve or even exceed the performance of the single quantum classifier.
	
\begin{figure}
		\subfigure[]{\includegraphics[scale = 0.45]{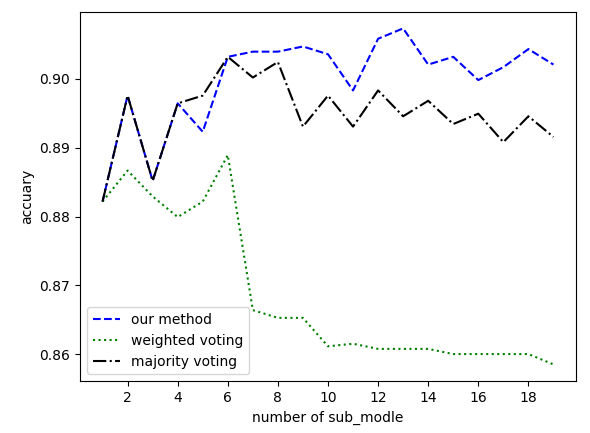}}
		\subfigure[]{\includegraphics[scale = 0.45]{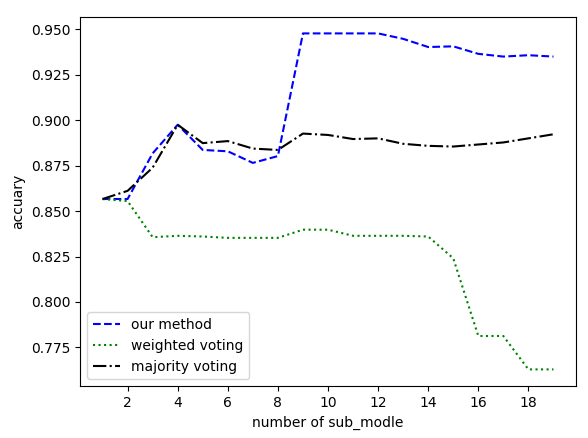}}
		\caption{The numberical simulation of combinated result. (a) The change of accuracy as the number of 4-qubit base learners increases. (b) The change of accuracy as the number of 6-qubit base learners increases.}
		\label{fig6}
\end{figure}

The combination strategy for the confusion matrix proposed in this paper is compared with the majority voting and the weighted voting with accuracy as the weight.
20 quantum base learners of 4-qubits are combined. As the number of quantum base learners increases, the accuracy of results changes with the number of learners, as shown in Tab. \ref{tab2} and Fig. \ref{fig6}(a).
	
Since the Bagging trains the quantum-based learner with data getting by random sampling, the base learner may express better classification ability for part of the dataset. Through experimental data, it is found that quantum base classifiers with good classification for the training subset can still maintain good performance when the whole training set is tested. 
QEL for 6-qubits base learner still show such characteristic. 
When 20 quantum base learners of 6-qubits are assembled, the change of accuracy as the number of base learners increases is shown in Fig. \ref{fig6}(b).	
\begin{table}[ht]
\caption{The comparison of variation of loss and accuracy with number of base learners and combination strategies.}
		\label{tab2}	
		\begin{tabular}{l|cccc}
			\hline
			\diagbox{Combination strategies}{Numbers of base learners}&5&10&15&20  \\
			\hline
			majority voting & 0.8821&0.8612&0.8608&0.8585\\
			weight voting with accuracy &\textbf{0.8975}&0.8976&0.8953&	0.8915\\
			weight voting with confusion matrix&0.8923&\textbf{0.9051}&\textbf{0.9043}&\textbf{0.9039}\\
			\hline	
		\end{tabular}
	\end{table}
	
Finally, in order to verify the stability of EL in this paper, 1000 samples are randomly sampled from MNIST digits dataset for testing the single learner and the ensemble learners respectively 10 times. Average accuracy is defined as $\overline{acc}=\frac{1}{n}\sum_{i=1}^{n}acc_i$. Variance is defined as $s_{acc}^{2}=\frac{1}{n}\sum_{i=1}^{10}(acc_i-\overline{acc})^2$, where $acc_i$ represents the accuracy of $i$-th prediction.
As shown in Tab. \ref{tab3}, when the accuracy of the single learner without ensemble is high, the performance of the Bagging method in EL is slightly improved. By analyzing the data in the experiment process, it is found that the reasons are mainly divided into two aspects. On the one hand, the accuracy of base learners in 6-qubits system is higher than 4-qubits system for binary classification. It may reduce the accuracy of the results when some learners without pruning participate in assembling. On the other hand, it is found that the difference between each quantum base learner is very small by comparing the differences between base learners. It means that the wrong prediction in the base learner can not be corrected with the addition of subsequent learners, but the error is ``consolidated" instead.
	\begin{table}[ht]
		\caption{Comparisons of average accuracy and variance of predictions with different combination strategies, where $\#Q$ and $\#P$ represent the number of qubits and the number of parameters in base learner, respectively. }
		\label{tab3}
		\begin{tabular}{lcccc}
			\hline
			\multirow{2}{*}{}&\multicolumn{2}{c}{$\#Q$ = 4, $\#P$= 12}&\multicolumn{2}{c}{$\#Q$ = 6, $\#P$= 81} \\ \cline{2-5}
			&$\overline{acc}$ & $s_{acc}^{2}$ & $\overline{acc}$ & $s_{acc}^{2}$\\
			\hline
			single
			learner&0.884&$8.000\times10^{-5}$&0.911&$9.036\times10^{-5}$ \\
			majority voting&0.851&$4.521\times10^{-5}$&0.762&$0.134\times10^{-3}$ \\
			weight voting with accuracy&0.897&$6.241\times10^{-5}$&0.904&$7.781\times10^{-5}$ \\
			weight voting with confusion matrix&\textbf{0.904}&$4.824\times10^{-5}$&\textbf{0.946}&\textbf{$3.461\times10^{-5}$} \\
			\hline
		\end{tabular}
	\end{table}
	
\section{Discussion \& Conclusion}

Recently, some progress has been made on QEL \cite{schuld2018quantum,macaluso2020quantum,macaluso2020quantum1,araujo2020quantum}. Tab. \ref{tab4} shows the comparison of our work with previous references. In Ref. \cite{schuld2018quantum}, the concept of quantum ensemble classifier is proposed, and preliminary theoretical proof and analysis are carried out. According to the method proposed in Refs. \cite{macaluso2020quantum,macaluso2020quantum1}, the qubits resources required by the model and the depth of the circuit will increase continuously with the increasing of the training sample size. Besides, their model can only classify one test sample at a time. In Ref. \cite{araujo2020quantum}, an optimized quantum ensemble classifier is proposed. However, their classier is difficult to execute on a quantum computer with the increasing of optimization steps.
While in our framework, the number of qubits and quantum circuits will not change with the increasing of the training sample size. Compared with the single learner model dealing with the whole training dataset, our QEL model can be applied to classification tasks of large-scale data with the same accuracy but a faster training process.	

\begin{table}[ht]
		\caption{Comparisons of our work with Refs. \cite{schuld2018quantum,macaluso2020quantum,macaluso2020quantum1,araujo2020quantum}.}
		\label{tab4}
		\begin{tabularx}{18cm}{XXXXX}
			\hline
			~&Ref. \cite{schuld2018quantum}&Refs. \cite{macaluso2020quantum,macaluso2020quantum1}&Ref. \cite{araujo2020quantum}& Ours   \\
			\hline
			Contributions & The concept of quantum ensemble classifier. & The untrained quantum ensemble classifier. & A trainable quantum ensemble classifier & A hybrid ensemble framework with a new combination strategy. \\
			\hline
			Limitations &  & Untrained quantum classifiers, one sample be tested at a time.& Huge quantum circuit with network increasing. & Requirements of memory.\\
			\hline
		\end{tabularx}
\end{table}

In addition, it should be noted that the accuracy of ensemble classifiers largely depends on the choice of quantum base learners.
The simulation results show that when the base learners reach a certain number, the performance of the model will not be improved
significantly. The reason lies in the fact that the diversity of base learners is insufficient.
How to obtain the quantum base classifier in difference as much as possible should still be the focus of QEL research, which is also suitable for other tasks of quantum base learners.

In conclusion, we have presented a hybrid  homogeneous QEL framework.
Meanwhile, the confusion matrix combination strategy that considers the performance and similarity of base learners is proposed. A trainable small-scale QCNN is applied as base learners in our framework for the classification task. Our simulation results show that the proposed strategy can achieve better accuracy in most cases than the majority voting strategy and the weighted voting strategy that uses accuracy as weight.

EL is an important method to improve model performance, but the nested learning framework mentioned in this paper still consumes a lot of computer memory. Frequent data exchange between the quantum computer and classical computer may not conducive to practical application in the future. Relevant research on QEL is still lacking, and its practical value still needs further explore.

\section*{Data availability statement}
The data that support the findings of this study are available from the corresponding author upon reasonable request.

\section*{Acknowledgements}
This project was supported by the National Natural Science Foundation of China (Grant No. 61601358).


\end{document}